# Models of the Mass-Ejection Histories of pre Planetary Nebulae.
# II. The Formation of the Butterfly and its Proboscis in M2–9

Bruce Balick[1], Adam Frank[2], Baowei Liu[2], Romano Corradi[3]

[1] Department of Astronomy, University of Washington, Seattle, WA 98195-1580, USA; balick@uw.edu

[2] Department of Physics and Astronomy, University of Rochester, Rochester, NY 14627, USA; afrank@pas.rochester.edu, baowei.liu@rochester.edu

[3] Gran Telescopio de Canarias, S/C de Tenerife, La Palma, E38712 Spain; romano.corradi@gtc.iac.es

**Abstract**

M2–9, or the "Butterfly Nebula", is one of the most iconic outflow sources from an evolved star. In this paper we present a hydrodynamic model of M2–9 in which the nebula is formed and shaped by a steady, low-density ("light"), mildly collimated "spray" of gas injected at 200 km s$^{-1}$ that interacts with a far denser, intrinsically simple pre-existing AGB wind has slowly formed all of the complex features within M2-9's lobes (including the knot pairs N3/S3 and N4/S4 at their respective leading edges, and the radial gradient of Doppler shifts within 20″ of the nucleus). We emphasize that the knot pairs are not ejected from the star but formed *in situ*. In addition, the observed radial speed of the knots is only indirectly related to the speed of the gas injected by the star. The model allows us to probe the early history of the wind geometry and lobe formation. We also formulate a new estimate of the nebular distance $D = 1.3$ kpc. The physical mechanism that accounts for the linear radial speed gradient in M2–9 applies generally to many other pre planetary nebulae whose hollow lobes exhibit similar gradients along their edges.

## 1. Introduction

Pre planetary nebulae ("prePNe") are nascent planetary nebulae ("PNe") that emerge from mass ejections during the ascent of the AGB branch and then evolve over the course of up to a thousand years (prior to the onset of ultraviolet ionization). As a group they have made for many of the most beguiling images ever obtained by the Hubble Space Telescope[1] (hereafter "HST"). In this paper we employ numerical hydrodynamic simulations to recreate the present structure of one of the most famous of all pPNe, the Butterfly Nebula M2–9, and to follow its evolution to within about a century from the onset of the winds that have shaped it.

The ultimate goal of hydrodynamic ("hydro") simulations of pPNe is to characterize the flow conditions that have shaped the nebulae. Physically realistic models are also a very powerful way to explore their early histories. In recent years we have been undertaking hydro studies of pPNe of varies morphologies, including CRL618 (Balick et al. 2013ApJ...772...20B; hereafter "B+13"), and OH231.8+04.2 (Balick et al. 2017ApJ...843..108B; hereafter "B+17").

M2–9—also known as Minkowski's Butterfly and the Twin Jet Nebula—was discovered in 1947

---

[1] This paper is partially based on observations made with the NASA/ESA Hubble Space Telescope, obtained from the MAST Archive at the Space Telescope Science Institute, which is operated by the Association of Universities for Research in Astronomy, Inc., under NASA contract NAS 5-26555. Support for MAST for non-HST data is provided by the NASA Office of Space Science via grant NAG5-7584 and by other grants and contracts.

by R. Minkowski. M2–9 is especially notable among its many siblings for the significant changes in its internal illumination and excitation on time scales of a century, as shown in the 12-year sequence of Hα and [OIII] images in Fig. 5 of Corradi et al., 2011A&A...529A..43C; hereafter "C+11"). The large-scale structure of M2–9 has an axial symmetry that is replicated very closely on opposite sides of the nucleus. (That is, the putative particle beams that rotate and locally illuminate the lobe edges do not appear to alter the shapes of the lobes' surfaces. Thus we shall ignore the dynamical influence of the beams in this study.) Accordingly, we shall perform two-dimensional simulations adopting persistent axisymmetric winds as the shaping agent for the strikingly regular, nested, and extended lobes of M2–9.

In section 2 we will present the steady-state flow simulations of the spatial present spatial structure of M2–9 (cf. Fig. 1). We shall find that this model successfully fits all of the observed features of M2–9, including the bulb, the sheath that surrounds the bulb, and the extension of the bulb called the feathers, the outer lobe, and the conspicuous knot pairs N3/S3 and N4/S4 at the tips of its inner and outer lobes (respectively). In section 3 we used the model to recreate the evolution of the lobes over time $t \leq 300$ y since the wind injection began. The long-slit kinematic observations of the patterns of Doppler shifts are used to evaluate the kinematic predictions of the model in section 4. In section 5 we assemble the most certain of the observations and model predictions to re-estimate the distance to M2–9. In the final section we shall summarize the major findings of the paper and explore how the model informs our understanding of the shaping and kinematics of other pPNe.

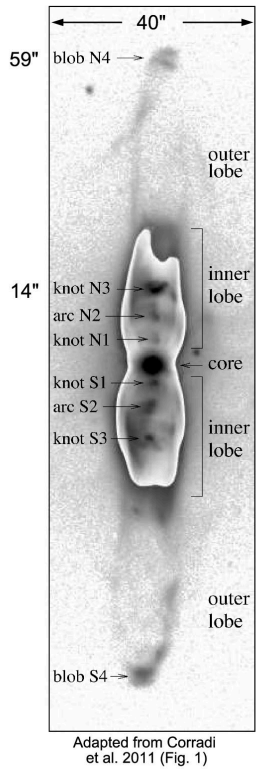 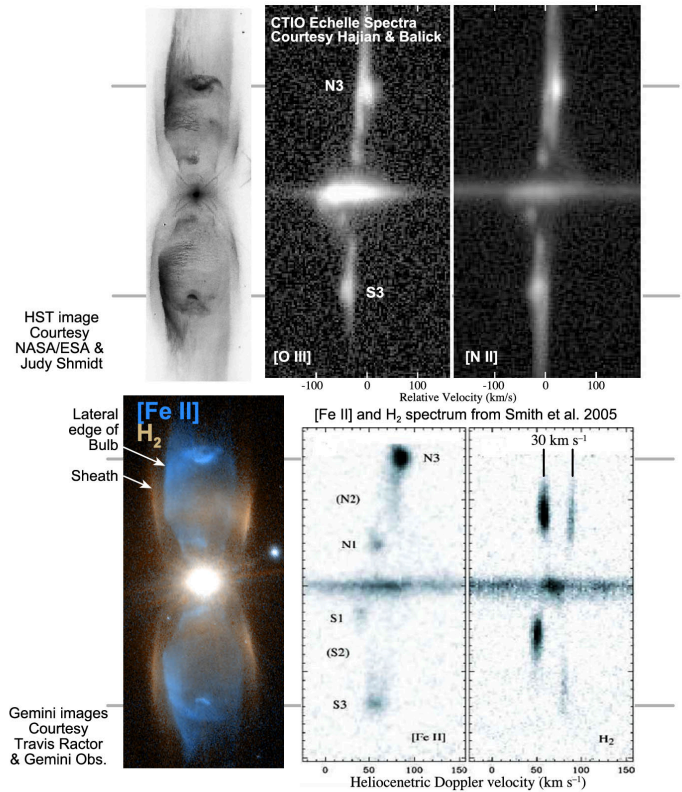

Fig. 1. A montage of the observations of M2–9 used throughout this paper. Key structural elements of the nebula are identified. Note the locations of knots N3/S3 and N4/S4 and the bulb and sheath.

**2. Models of the Current Morphology of M2–9**

2.1 M2–9 Today

M2–9's is a poster child for all types of bipolar nebulae created by symmetric pairs of flows from a central source. The observations that serve as the foundation of our simulations are compiled in Fig. 1. (Please refer to this figure in subsequent discussions.) Special spatial features of M2–9 that constrain the models are described in section 2.1. The tools of our hydro modeling are discussed in section 2.2. The outcomes of models that match the current structure of M2–9 are presented in sections 2.3 and 2.4.

2.1.1 Knots

Various very conspicuous knots lie on or near the symmetry axis (Fig. 1). Knot pairs N1/S1 and N2/S2 have been moving steadily from the east to the west edges of the bulb (or vice-versa) since 1947 with a possible period of 90 years (see C+11). Close inspection of images over the past 30 y shows no trace of any bulge or other dynamical distortion where the beam strikes the bulb walls. The knots N1/S1 and N2/S2 seem to be the result of a beam illumination pattern of no dynamical significance that is irrelevant to this study. These knots are not considered further.

Knot pairs N3/S3 are located near the tips of the inner lobes. They are bright and seem flattened in HST images. Our analysis of archival HST images in 1997 and 2001 suggest an upper limit to their proper motions is of order 50 km s$^{-1}$ at an assumed distance of 1.3 kpc (C+11). However, both N3 and S3 varied in shape somewhat during this time rendering proper motion studies difficult.

As shown by C+11, knot pairs N4/S4 are presently separated from each other by an angle $\theta$ = 115″ with proper motions d$\theta$/dt = 0″.046 ± 0″.004 y$^{-1}$. Thus, if d$\theta$/dt has been constant—a presumption that we shall challenge later—then the kinematic ages of these knots are $\theta$/(d$\theta$/dt) = 2500 y independent of nebular distance and inclination. C+11 and Smith et al. (2005 AJ....130..853S; hereafter "S+05") suggested that the age of 2500 y applies to all of M2–9's features, though Solf (2000A&A...354..674S) and Clyne et al. (2015A&A...582A..60C, hereafter C+15) take exception.

2.1.2 Large-scale Features

Aside from N1/S1 and N2/S2 the large-scale structure of M2–9 appears strikingly axisymmetric. The "inner lobes" consist of a mirror-image pair of co-axial features of similar lengths, the thinner (wider) of which is called the "bulb" ("sheath"). The side edges of the bulb and sheath are separated by a resolvable gap. The leading edges of the bulbs lie just beyond knots N3 and S3.

The bulbs appear to be hollow in all high-quality images in which they are visible, most prominently in various emission-lines. Very faint dust-scattered starlight is also seen from their edges through the line-free filters F547M and F814W in archival HST images. The bulb interiors may be filled by sparse gas that does not emit radio, infrared, or optical light (no x-ray images of M2–9 are available). The pressure of interior gas is likely to sustain the thin walls of the bulbs.

Deeper emission-line images show "feathers", or smooth vertically striated emission-line extensions beyond knots N3/S3 and the closed bulb tips (C+11 and C+15). The feathers seem to lie within an open cylinder extending outward from the outer edges of the bulb on the nebular symmetry axis. They fade rapidly with distance from the nucleus. Ratios of aligned emission-line images indicate that the bulbs and feathers have very similar spectra. Note that they are also visible in the images of the shock tracer [Fe II]$\lambda 1.64\mu$m (Fig. 1) made using adaptive optics and obtained from the archives of the Gemini Observatory[2] (courtesy Travis Rector).

The outer edge of the sheath is seen exclusively in dust-scattered starlight as well as the Gemini $H_2$ S(1–0) $\lambda 2.12\mu$m image. The gap between the bulbs and sheath appears to be largely empty in high-resolution images. No emitting gas has been detected beyond the sheath, including $^{12}$CO (Castro-Carrizo et al., 2017) so there is no direct evidence of an external medium. However, our hydro models of M2–9 require such an external medium (presumably the slow winds from an AGB star) in order to confine and shape the inner and outer lobes as they evolve.

C+11 suggest that both [Fe II] and $H_2$ lines are excited locally by shocks throughout M2–9 (see also S+05). The spatial dissimilarities of the [Fe II] and $H_2$ emission from the bulbs and sheaths are very clear in the Gemini images of Fig. 1. Presumably each line is collisionally excited in shocks of very different locations and speeds. [Fe II] lines are usually seen, among other places, in shocks within supernova remnants and very massive stars with fast winds. In contrast, $H_2$ lines arise along the lobe edges of bipolar PNe and nebular bullets such as OMC-1 where inflowing gas creates shocks at high obliquity (that is, shock speeds $\leq 50$ km s$^{-1}$).

2.2  Model Methodology

Hydro simulations were made using AstroBEAR, an adaptive-mesh-refinement code that uses a Riemann solver to follow the varying Eulerian state variables for various types of flows over time (Carroll-Nellenback et al. 2013, Cunningham et al., 2009, ApJS, 182, 519). The spatial resolution of the mesh adjusts as needed wherever pressures change rapidly. Here we mention a few highlights of the code. See the more detailed description of AstroBEAR in B+13 and B+17.

Our simulations are restricted to a 2-dimensional slice through the nebula that contains the symmetry axis. In previous papers (B+13 and B+17) we noted that a comparison of 2-d and 3-d models showed only slight changes on model outcomes unless substantial thin-shell or shear instabilities are present. Such instabilities are so disruptive to the low-density outer structures of M2–9 that the observed knots N4/S4 cannot form. Moreover, each 3-d model of adequate spatial resolution requires prohibitive amounts of computer time. We did not attempt such models.

The 2-d computational grid is $32 \times 128$ kau on a side with the outflow nozzle at the origin and the nebular symmetry axis along the $y$-axis. The grid is filled with static, stratified ambient gas at time $t = 0$, where "ambient gas" refers to the surrounding pre-flow AGB wind whose density distribution is presumed to decline as $1/r^m$ (where $1.8 \leq m \leq 2.3$) and whose speed is negligible compared to that of the gas injected at the nozzle. The cell size is 500au/$2^n$, where $n$ is an integer $\leq 6$ determined by an adaptive grid algorithm. Radiative cooling is computed from the coronal cooling rates of Dalgarno & McCray (1972).

Each 2-d run required 2 to 24 h to complete. So, in order to enhance computational speed, $n$ is set to 6 for the first 300 y, 5 for the following 900 y, and 4 thereafter. Thus the maximum resolution of the simulations is matched to the scale of the growing nebular structure. The time steps in the model are on the order of days or weeks. Synthetic images of the hydro state variables are created every 50 y and compiled in large output files for display using standard display tools. It is important to note that AstroBEAR does not generate emission-line images or synthetic

---

[2] The Gemini Observatory is managed by the Association of Universities for Research in Astronomy under a Cooperative Agreement with the National Science Foundation.

position-velocity diagrams for comparison to observations of speed patterns.

The grid is initially filled with a stationary "ambient" AGB wind of the form $n_{amb}(r) = n_{amb}(r_0)$ $(r_0/r)^m$ (cm$^{-3}$), where $r_0$ is the radius of the spherical nozzle, $n_{amb}(r_0)$ is the AGB wind at the nozzle, and $m$ is a free parameter. A steady tapered conical jet, or "spray", flows radially from the boundary of a spherical nozzle starting at time $t = 0$. The radius of the nozzle is 500 au. The spray at the nozzle boundary is characterized at all times by its hydro parameters at the nozzle, $n_{spray}(r_0)$, speed $v_{spray}(r_0)$, temperature $T_{spray}(r_0)$, and its geometry $\phi$. All of these are fixed throughout the run. $\phi$ is the 1/e width of a Gaussian that is used to modulate the polar distribution of $n_{spray}(r,\phi)$ and $v_{spray}(r,\phi)$ at all $r$. This tapered momentum flow means that the momentum of the spray, proportional to $n_{spray}(r,\phi) v_{spray}(r,\phi)$, is concentrated along its symmetry axis. Note that we adopt spatial units of kilo au ("kau") since these are the natural units of the computations. 1 kau $\approx 0.06$ pc $\approx 1.5 \times 10^{11}$ km. The units of flow speed, density, and other quantities are standard.

For reasons made clear in section 2.3, we adopt a very "light" spray in which $n_{spray}(r_0) \ll n_{amb}(r_0)$. Thus the low-density spray (like a steady puff) interacts strongly with the ambient AGB wind right from the outset of spray injection. Lobes slowly form and grow over the course of 2500 y. Their shapes and dimensions mimic those of the present inner lobes of M2–9 when the initial conditions of the ambient gas and the flow are properly chosen (section 2.4).

2.3 Best Simulation of the Inner Spatial Structure of M2–9

The starting point for the simulations is this: a spray exits the nozzle with mass density $n_{spray}(r_0) \ll n_{amb}(r_0)$, highly supersonic speed (Mach number $\gtrsim 100$), and opening taper angle $\approx 40°$ and encounters a much denser and static AGB wind near the nozzle. The best values of $n_{amb}(r_0)$, $n_{spray}(r_0)$, $v_{spray}(r_0)$, $m$, and $\phi$ are found from the model that best fits the observations at $t = 2500$ y.

We ran over 70 simulations before finding the most satisfactory (albeit imperfect) fit shown in Fig. 2. The most successful of the models at $t = 2500$y is shown in Fig. 2. The locations of few key features of the model are identified in the panels. In this model $r_0$ is 0.5 kau, $n_{amb}(r_0) = 1 \times 10^4$ cm$^{-3}$, and the descriptors of the injected gas are $n_{spray}(r_0) = 400$ cm$^{-3}$, $v_{spray}(r_0) = 200$ km s$^{-1}$, $T_{spray}(r_0) = 100$ K, and $\phi = 40°$. The corresponding mass flow rate through the nozzle is $\dot{M} \approx 10^{-8}$ M$_\odot$ y$^{-1}$, the total injected mass is $2 \times 10^{-5}$ M$_\odot$ after 2500 y, and the total injected momentum is 0.004 M$_\odot$ km s$^{-1}$ ($10^{36}$ gm cm s$^{-1}$).

Fig. 2a shows the density distribution at $t = 2500$ y. The inner and outer lobes are found in the lower and upper halves of this panel, respectively. Overall, the model nicely reflects the shape of M2–9 seen presently. The low-density interior of the bulb is filled by low-density streamlines of injected gas from the nozzle. The expected rim of ambient gas that they have displaced is found at its outer perimeter. Dense clumps at $y = 20$ (45) kau represent knot pairs N3/S3 and N4/S4, respectively; however, knots N4/S4 are only twice as far from the nucleus as N3/S3, not a factor of four as observed. The sheath has the appropriate width; however, its interior density is uniform, not edge brightened, as images suggest. (Other model shortcomings are discussed later.) Finally, we find that increasing the value of $r_0$ by more than 50% smears the spray (and increases the momentum injection rate) such that N4/S4 cannot form.

The streamlines contact the rim of the inner lobes at a reverse shock along the edges of the bulb (see lime-green line in Fig. 2b). The normal component of their momentum is transmitted into the gas. The local gas temperatures and speeds change abruptly across this shock (Figs. 2b, c, d).

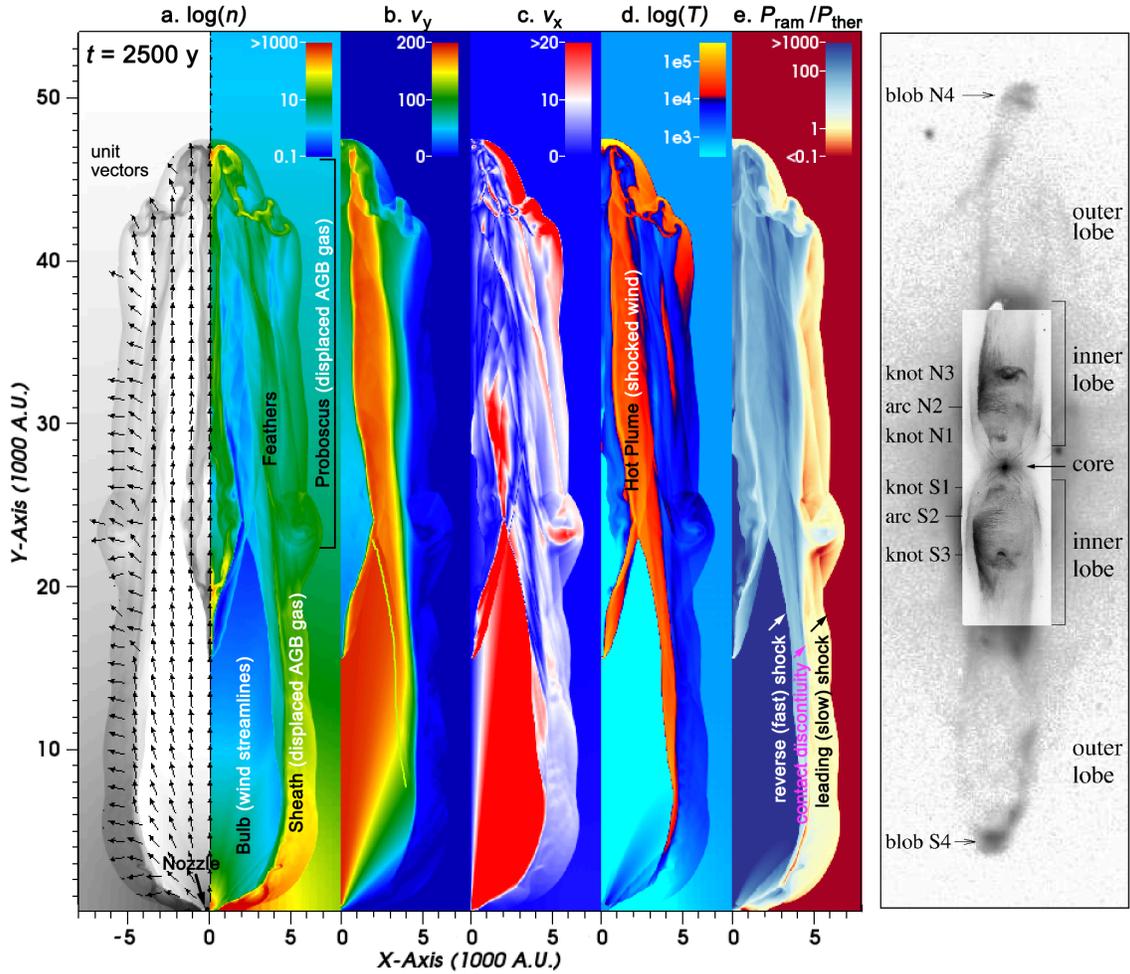

Fig. 2. Outcomes of the best simulation after 2500 y and selected observed images of M2–9. Panels from left to right:
  a. The density distribution (using two color schemes). Unit flow vectors are shown on the left side.
  Note the locations of N3/S3 at $y \approx 18$ kau and N4/S4 at $y \approx 45$ kau.
  b. The $y$-component of the speed pattern.
  c. The $x$-component of the speed pattern. Note the changes in the legend from panel b.
  d. The temperature distribution. Note the location of the hot plume.
  e. The ratio of ram to thermal pressure. Not the locations of the shocks and the contact discontinuity.

The shock speed of the inner shock is of the order of the flow speed at the point of contact ($\leq 200$ km s$^{-1}$) times the cosine of the polar angle of the streamline. Thus the speed of the inner shock varies steadily along its length. Nonetheless, the shock speeds are sufficiently large to excite the bright tracers in the post-shock zone such as [Fe II], [S II], and [N II].

The outer edges of the sheath expand horizontally (Fig. 2c) and supersonically into the cold ambient gas, preceded by a leading shock in which H$_2$ is excited. The locus of this shock is indicated in Fig. 2e. A contact discontinuity ("CD") lies between them pair of shocks. There is no pressure gradient across the CD to force gas through it. On the other hand, over time the embedded CD drifts outward ($x$-direction) into ambient gas near the outer edge of the sheath. Gas formerly upstream of the CD intrudes through the CD. As we discuss in section 4, this slow but steady intrusion of ambient gas leads to complex speed gradients between the CD and the inner shock. This is illustrated to the immediate right of the lime-green contour Fig. 2b.

The bulb and the surrounding sheath are approximate analogues of the wind-heated bubble and rim in the inner zones of classical round and elliptical PNe, respectively; see Kwok, Purton, &

Fitzgerald, 1978ApJ...219L.125K and Toalá & Arthur, 2014MNRAS.443.3486T. The latter paper shows that the radial streamlines can induce highly disruptive thin-shell instabilities along the inner shock rim at the bulb's surface. No such instabilities are observed in M2–9. In practice this places an upper limit on the flow injection speed at 225 km s$^{-1}$. The minimum injection speed, 180 km s$^{-1}$, is set by the requirement that the knots reach their observed positions.

2.4 Simulation of the Proboscis of M2–9

The formation of the outer lobe, or the proboscis, of M2–9 is a very different story from that of the inner lobes. The proboscis is powered by fast, injected gas that has bypassed N3/S3 and re-converges further upstream to form N4/S4. This bypass flow traverses the inner shock where it is heated to >10$^4$ K as it starts to converge to the $y$-axis. An invisible, low-density, high-speed "hot plume" forms. The plume is easily seen above N3/S3 in Figs 2b and 2d.

The faint shock along the outside edge of the hot plume forms shocked vertical "feathers" where conditions resemble those in the shock on the inside edge of the bulb. Morphologically, the feathers appear to be outward extensions of the bulb. Line ratios of [N II], [S II], and [Fe II] in the bulb and feathers mimic one another. The vertical striations along the surface of the feathers appear to be the result of surface shears that we cannot simulate on a 2-D grid. Beyond the feathers the proboscis is seen only in starlight that was scattered from the displaced ambient gas and dust.

The knots N4/S4 contain a mixture of gas injected gas delivered by the plume and displaced ambient gas. In our simulations the fraction of fast gas in N4/S4 is about 60%. So the speeds of N4/S4 are on the order of 120 km s$^{-1}$ whereas observations of N4/S4 by C+11 show a faster speed. This fraction (and the knot speed) steadily increases as the downstream density of slow ambient gas declines over time (see section 4).

As noted above, the most glaring shortcoming of the hydro model on M2–9 is that N4/S4 are not four times as far from the core as N3/S3. We explored remedial initial conditions without any success. For example, increasing the flow speed from 200 to 225 or 250 km s$^{-1}$ resulted in highly disruptive instabilities throughout the proboscis beyond N3/S3. As a consequence, the hot plumes do not develop and knots N4/S4 can't form. We reran the "baseline" simulation of Fig.2 with the same ambient density structure except that the decrease with radius scales as $(r_o/r)^{2.2}$. This had no impact on the location of N4/S4 at 2500y; however, N3/S3 were further from the nucleus relative to the baseline result and the outer lobes were too narrow. Decreasing $n$ to 1.8 or less led to an open proboscis in which N4/S4 cannot form.

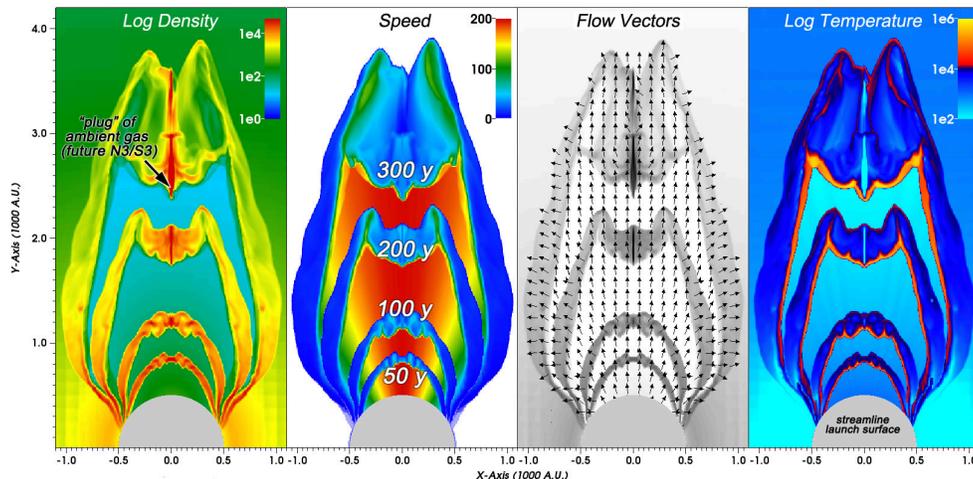

Fig 3. The structure of M2–9 in its first 300 y. The flow density, outflow speed, flow patterns (unit vectors), and the temperature are shown in panels, a, b, c, d, respectively at $t$ = 50, 100, 200, and 300y. The nozzle is shown in gray.

## 3. The Early Years

The rapidly declining density of the external medium suggests that the expansion pattern will be essentially ballistic once the density of the growing rims exceeds that of the local AGB wind. In the present model the inner and outer lobes are largely ballistic by $t$ = 500 y. Therefore the basic structural elements of M2–9's present structure developed quickly during its infancy.

Frames from the model from 50 through 300y are compiled in Fig. 3. The cell width used in these simulations is 500 au/$2^6 \approx$ 8 au. To understand the slow expansion speeds of the compressed rim it is useful to recall that the density of the injected spray is 25 times less than that of the stationary ambient gas at the tip of the nozzle—and even smaller at other polar angles owing to the taper.

Note the dense "plugs" that form where streamlines are deflected towards the $y$-axis by the inertial pressure of the confining ambient gas. (Some of their details may be artificial consequences of the nozzle pixelation and the approximate cooling curve adopted for the simulations; however these plugs form in all simulations for $n$ = 4, 5, or 6.) The slight asymmetry in the left and right halves of the panels is a result of the pixelated rendering of the nozzle surface used in the computations. The differences are insignificant.

Almost immediately the cold dense rim of freshly swept up ambient gas develops where the streamlines end (Fig. 3). By $t$ = 100 y there is a net flow of gas moving along the rim and upwards towards its leading tip. Most of the gas at the tip is slow ambient gas. It cools efficiently, so a compact and cold plug develops. Its increasing mass and declining speed are the result of ambient gas of low specific momentum that has been incorporated within the plug.

At $t$ = 200 y—while the density of the inner ambient medium and its rate of incorporation into the plug are greatest—the tip of the flow traverses only 1800 au. It average speed is 45 km s$^{-1}$, or about one-fourth of the flow speed at the nozzle. The dense plug appears to survive without being "crushed" by the much faster gas that overtakes it (B+17). In fact, the plug slows down further and becomes an obstruction for the streamlines that reach it. The streamlines push around the plug forming a pair of small horns. By $t$ = 300y the evolving rim has assumed the embryonic form of the knots and lobes seen at later times. The leading plug has slowed to 35 km s$^{-1}$. It maintains a speed of $\approx$30 km s$^{-1}$ and soon morphs into N3/S3.

Note that the sheath forms early from a pressure wave that precedes the slowing rim into the ambient gas. Many of the other final characteristics of M2–9 are also in an incipient stage by 300 y. The horns of knot-deflected gas later become the hot plumes that fashion the proboscis. Thus the formation of the outer lobes is essentially inevitable as the hot plume is tilts towards the symmetry axis, guided by the pressures of the stagnation pocket of N3/S3 and static ambient gas further downstream. If these flows were any denser (heavier) as they enter the outer lobe they would diverge away from the symmetry axis and, so, N4/S4 could not form.

**4. Nebular Kinematics**

None of the kinematic observations shown in Fig. 1b were taken into consideration when constructing the best model (Fig. 2). Thus these data are useful for evaluating various predictions of the model. In this section we shall review the highlights of the kinematic observations and, subsequently, compare them to the predictions of the hydro simulations.

4.1 Current Kinematics

The most applicable kinematic data consist of the long-slit observations shown in Fig. 1b. The optical spectra of [N II] and [O III] were obtained by Hajian & Balick (private communication) using the Echelle Spectrograph on the 4-m Blanco Telescope at Cerro Tololo Interamerican Observatory ("CTIO")[3]. The observations were made in 1.″5 seeing during July 1998 though a 1″-wide slit oriented north-south. The spectra have spectral resolution of 7 km s$^{-1}$. The near IR spectra of [Fe II] and $H_2$ were obtained 15 months earlier by S+05 at CTIO with about the same spectral resolution. See that paper for technical information.

Ignoring N1/S1 and N2/S2, the emission-line spectra north and south of the nucleus are mirror images of one another except that the northern lines are slightly redshifted and the south lines are slightly blueshifted with respect to the average Doppler shift of the broad nuclear emission lines. Thus the south lobe is approaching. See Solf (2000) and S+05 for a much more detailed description of the large-scale Doppler patterns in the optical and infrared spectra, respectively.

Recall that the [Fe II], [N II], and [O III] lines trace the locus of the inner shock at the edge of the bulb. Their respective kinematic patterns should be (and are) very similar. All of the long-slit spectra show a small but clearly positive overall gradient in their Doppler shifts in the $y$-direction, $dv_y/dr$, from the base of the bulbs through N3/S3 and the feathers. This is the "kinematic gradient" whose observed magnitude is 10 km s$^{-1}$ from the core to N3/S3 before any inclination corrections (a factor of 3.2). In addition to this "vertical" speed gradient in the $y$-direction, the lobes expand "horizontally" in the $x$-direction at $v_x \approx$ about ±15 km s$^{-1}$ with very little gradient along the slit. These results agree very well with the heuristic inclination-corrected fit to Echelle data by C+15.

In contrast, the $H_2$ lines seen through the slit on each side of the nucleus from the front and rear edges of the sheath (Fig. 1) at Doppler shifts of ±15 km s along the line of sight with no kinematic gradient (i.e., no overall spectral tilt). No mass moves vertically along the edge of the sheath as it slowly opens horizontally. The models show that the gas near the $H_2$ emission region is entirely ambient gas that was initially pushed aside by the tips of the outer lobes as they moved vertically. The present motions of the outer edge of the sheath are a perpetuation of the same horizontal flow pattern because the nearby ambient gas in their path is unable to slow the flow.

---

[3] The CTIO is managed by the Association of Universities for Research in Astronomy under a Cooperative Agreement with the National Science Foundation.

4.2 Dynamics of the Inner Lobes.

The most obvious feature of the long-slit spectra of [Fe II], [N II], and [O III] is the $y$-gradient in speed from the nucleus through the tips of the feathers (Fig. 1b). The positive vertical speed gradient arises as follows. As noted earlier, the streamlines strike the rim and the inner shock obliquely. The locally transverse component of momentum is unaffected by the impact. Given the lobe's flame-shaped geometry, the post-shock flow speeds will be larger at smaller polar angles (near the top of the lobe) —where the flow vectors are nearly parallel to the shock—than at the base. Accordingly, there is an easily discernible positive gradient in the vertical speed of the gas where the emission lines of [Fe II], [N II], and [O III] arise (adjacent to the lime-green line in Fig 2b). (This explanation for the vertical speed gradient along the walls is robust for most lobe geometries; e.g., Lee & Sahai, 2003ApJ...586..319L, hereafter "LS03".)

As noted in section 2.2, the slow horizontal drift of the CD allows slow ambient gas to intrude through the CD whence it mixes with faster gas that passed through the inner shock earlier. This causes a sharp speed gradient between the inner shock and the CD. The duration of this penetration of ambient gas is longer at the (older) base of the sheath than at the top. The outcome is that the region between the inner shock and the CD is a region of complex yet systematic kinematics with substantial speed gradients in the $x$ and $y$ directions.

For several technical reasons we are unable to make a detailed quantitative comparison of the kinematic data and predictions of the model. Firstly, as noted earlier, AstroBEAR does not compute local line emissivities. Secondly, the visualization program that we used, VisIt, does not accommodate the rendering of a position-velocity diagram for 3-d volumes constructed from 2-d models. (This limitation will be fixed in the near future.)

However, VisIt can display a trace of the hydro state variables along a single straight "data readout line" that can be placed interactively on the panels of model outcomes, including those presented in Fig. 2. The value this tool is limited since the locus of the line tool is straight and inner shock is curved. Even so, it is clear that the pattern of flow speeds just beyond the lime-green line in Fig. 2b—the locus of the inner shock—is dominated by a gradient in $v_y$ of the appropriate slope. On the other hand, the absolute values of the speeds vary dramatically when the data readout line is translated only slightly. The net result is that the tilts of the lines in Fig. 1b are easy to understand qualitatively but tricky to replicate in detail. We are forced to accept a qualified victory.

The model successfully predicts that the sheath is steadily widening (Figs. 2a and 2c). However, the predicted speed of the sheath, $v_x \approx 8$ km s$^{-1}$, is half of the observed value. This discrepancy is stubbornly difficult to resolve in our simulations without degrading the fit of the model elsewhere. For example (as noted earlier), a increase of the injection speed at the nozzle results in unacceptably disruptive instabilities result throughout the entire proboscis.

4.3 Dynamics of the Outer Lobes

Recall that N3 and S3 are not launched from the core but rather originate in the plugs that develop with the rims of the lobes at very early times. These plugs quickly become much denser than the gas through which they move, so the motions of N3 and S3 are soon ballistic. Their speeds can be read directly from Fig. 2b or, better still, derived from the knots' positions measured directly from the AstroBEAR density panels in 100-year intervals starting a $t = 300$y. We have applied both procedures. The results of the two methods are consistent: N3 and S3 have steady vertical speeds of 30 km s$^{-1}$, in excellent agreement with their current values (C+15).

Spectra of [Fe II], [N II], and [O III] show that N3/S3 have slightly more extreme Doppler shifts than the edges of the bulbs in the same line of sight. The small "anomalous" component of knot velocity is on the order of 15 km s$^{-1}$ after deprojection. The anomalous speeds are not predicted by the model. We can only speculate on a cause. A series of almost annual images of M2–9 published by C+11 show clearly that N3/S3 are immersed within the collimated and precessing "lighthouse" jet whose flow speed exceeds $10^4$ km s$^{-1}$. Perhaps the denser knots N3/S3 deflect the beams so that the post-deflection speeds of the locally shocked gas are anomalous[4].

The present model tells an unusual story about the historical movements of N4/S4. The knots are merely ephemeral patterns of high density at the tips of a complex flow for the first 2000 y. During this time fast gas arriving through the converging hot plumes continuously roils the gas at the leading tip of the outer lobe. Blobs of compressed gas form and then dissolve, only to be occasionally replaced by new leading blobs. The tips of the outer lobes advance in occasional lurches. The average speed of the tips is about 100 km s$^{-1}$. After 2000 y the last of the blobs becomes a long-lived entity embedded in hot swirling gas. Its terminal speed is 120 km s$^{-1}$; i.e., ≈20% larger than its average speed.

**5. The Distance to M2–9**

C+11 measured the relative locations of N4 and S4 from archive images. They also measured an expansion age $\theta/(d\theta/dt)$ of 2500 y (to an accuracy of 10%), where $\theta$ is the angular separation between N4 and S4, and the true separation speed of these blobs. From this they calculated a distance $D = 1.3 \pm 0.2$ kpc to M2–9.

We present somewhat different approach for finding the distance to M2–9 using the model presented here in combination with selected observational data with minimal uncertainties. Suppose that we assume that N3/S3 lie near the forward edges of their bulbs (as the images show and the model requires), that the forward speed of the bulbs is 30 km s$^{-1}$ (as C+15 measured and the model suggests), that the N3-S3 separation speed has been constant (as the model suggests), and that the nebular age is 2500 y. Then these knots have traversed ±15 kau in 2500y. They now lie ±14" from the nucleus, so the distance to M2–9 is $D = 1.1$ kpc.

According to the model the present vertical speed of N4/S4 is ≈20% larger than its average value. If we adopt a correspondingly larger expansion age and repeat the estimates above, then knots N3 and S3 have now traversed 18 kau (in better agreement with Fig. 2a). The derived distance increases to $D = 1.3$ kpc, the same value found by C+11.

It is clear that a directly measured distance to M2–9 would be very valuable. A more precise expansion age using the knots N3/S3 is possible now simply measuring their proper motions; i.e., obtaining a new HST image of M2–9 in order to derive the proper motions of the knots since 1997. If their space motions are 30 km s$^{-1}$ and if $D = 1.3$ kpc their angular separations will have increased by 0".09 over 20 y. Such proper motions are easily measurable.

---

[4] If this is true, then we can expect that the anomalous Doppler shifts of N3 and S3 (relative to the trend line) will display equal but opposite speed changes that correlate in time with the changing longitude of the precessing beam. It will be interesting to see whether this is the case.

## 6. Summary and Conclusions

We have described a simple steady-flow hydro model that accounts for major structures, kinematics of M2–9 (Fig. 1) at the current time. In summary, spray-like winds from the nucleus are injected through a small nozzle into an ambient medium like that expected from the winds of an AGB star. The spray has an 1/e-angular width of 40˚, a polar speed of 200 km s$^{-1}$, and a density 1/25 that of the ambient gas that it displaces at the nozzle. We follow the shaping of the lobes of M2–9 for 2500 y (Fig. 2), its putative expansion age.

The model closely (but not exactly) replicates the myriad and complex features, shapes, and dimensions of M2–9, the shocked nature of the emerging spectra of from its inner lobes, the present locations and proper motions of knot pairs N3/S3 and N4/S4, the patterns of Doppler shifts found in medium-to-high-dispersion long-slit studies, and the largely complementary morphologies of [Fe II] and $H_2$ images (S+05). Furthermore, we found a very natural and robust explanation for the vertical speed gradient seen in most emission lines along the nebular symmetry axis as well as the constant speed and absence of any speed gradient seen in lines of $H_2$. Results of the model together with extant data have allowed us to re-estimate its distance with a result of $D$ between 1.1 and 1.3 kpc.

At the end of this section we will discuss the robust features of the model that provide significant insight into the physics of the speed gradients and related kinematic patterns seen observed in other pPNe.

6.1 The Historical Evolution of M2–9

A highly scientific value of the model is that it allows us to trace the evolution of the various structures of M2–9 back to within 100 years of the onset of spray injection (section 3). The process of forming the unusually complex structure of M2–9 is actually quite straightforward. Geometrically, the low-density spray burrows into a far slower AGB wind of far higher density and immediately inflates the nascent hollow bulbs. A thin rim of slow displaced and compressed ambient gas immediately forms along the leading edge of the bulb along with a dense leading "plug" of compressed ambient gas at its tip. The pair of plugs (analogous to the water wave at the prow of a speeding boat) become N3/S3.

The slow plug is soon overtaken and bypassed by the ongoing spray. Thus begins the process of forming the proboscis and knots N4/S4. The plugs immediately become and remain an obstacle within the much faster spray in which they are immersed. Under suitable conditions the bypass flow converges back to the symmetry axis downstream from N3/S3 where ephemeral blobs such as N4/S4 can form. At the same time, the gas initially displaced in the $x$-direction by the expanding (but decelerating) edges of the bulb forms a mildly compressed sheath of slow ambient gas. Both the bulb and sheath expand ballistically and supersonically.

The interaction of the streamlines with the rim of displaced gas creates shocks of speed ≲200 km s$^{-1}$ along the perimeter of the bulb. These shocks excite various emission line tracers. The obliquity of the impact and the effective shock speed vary somewhat with height (that is polar angle) along the curved walls of the bulb. Thus the emission lines exhibit a systematic gradient in Doppler shift when observed through a slit placed along the symmetry axis.

Once formed, the plugs move ballistically at 30 km s$^{-1}$ as they morph into N3/S3. The off-axis streamlines flow around the plug, creating a pair high-speed 'plumes' on a two dimensional grid (Figs. 2b and 2d). The plumes contain fast hot shocked gas that is constrained by the inertial pressure of the static gas that they displace. They sweep around the plug and, owing to the confining pressure of the ambient gas near the core, slowly converge towards the symmetry axis. (The model also predicts a thin axial column of cold gas released from N3/S3 that ascends towards and supplies mass to N4/S4. This column is not observed.)

After 2500 y relatively dense and stable pair of blobs (N4/S4) has formed at the outermost tip of the converging hot plumes (Fig. 2a). Each knot consists of a 40:60 mixture of swept-up ambient and fast injected gas, so its vertical speed is ≈120 km s$^{-1}$. The speeds of the blobs can steadily increase as the density of ambient gas declines and as high-speed injected gas is delivered from behind. At 2500 y the forward speed of N4/S4 is 20% larger than the average growth speed of the tip of the outer lobe. A larger fraction of injected gas would increase the distance and speed of N4/S4 from the nucleus.

The values of the initial conditions adopted for the model in Fig. 2 were obviously adjusted to provide a good fit to the extant spatial observations of M2–9. New observations (such as proper motions) are needed to test the model's predictions of its evolution. In principle, images in soft X-rays would also be useful, but their feasibility is a formidable challenge. ALMA images of the proboscis in CO or the cold dust distribution would also be helpful for evaluating the model.

6.2 Shortcomings

The present model, like all models, is a physics-based yet still hypothetical story of the evolution of M2–9. Any model of a nebula as complex as M2–9 has its limitations, and ours is no exception. To begin with there are eight free, non-orthogonal parameters that describe the nozzle, the ambient gas, and the injected spray that need to be adjusted for a good fit, somewhat rendering the optimization of the model a matter of informed guesswork. More free parameters might improve the quality of the model—at the expense of the complexity of running simulations.

The shapes of the lobes, the width of the sheath, the convergence of the plumes to form blobs, and the locations of the knots demand very specific combinations of $r_0$, $n_{amb}(r_0)$, $n_{spray}(r_0)$, and $v_{spray}(r_0)$. While we cannot be certain that our selection of initial conditions is optimum. Tests show that our parameter choices cannot be perturbed by more than a factors ranging from 10 to 50% and still yield satisfactory results. That makes the selected parameters robust choices. However, our model may not be the best possible.

There are many other limitations of the present model. The geometric structure of the "spray" and the ambient gas are obviously idealizations. Also, the ambient gas was presumed to be static for reasons of convenience. (Tests show that employing AGB winds speeds of 15 km s$^{-1}$ has little effect on the outcomes at the cost of hugely increased computational time.) We did not consider magnetic fields nor did we attempt any three-dimensional simulations. AstroBEAR has its own limitations; for example, it uses a very simplified, numerically efficient cooling curve that is vital for computing the structures of shocks and the formation of dense knots. Finally, we were unable to synthesize P-V diagrams from models for direct comparison to the kinematic observations.

In addition we find some troublesome outcomes of the simulations. Foremost among these are the precise locations of the knots N3/S3 and especially N4/S4. The latter are observed to be considerably further (by twice) from N3/S3 than predicted. Other shortcomings are these:
- The speed of the edges of the sheath, $v_s$, is observed to be 50% larger than predicted.
- The model predicts that the zone between the bulbs and the sheath is filled with ambient gas at densities ranging from 10 to 100 cm$^{-3}$ (Fig. 2a). Yet the images of M2–9 in Fig. 1 show that this zone is really a gap. A filled sheath should have been visible in reflected starlight in F547M and F814W images in the HST archives (N.B.: this concern does not apply to lobe edges seen in shock-excited $H_2$).
- Any major change in $n_{amb}(r_0)$ from $10^4$ cm$^{-3}$ yields a poor fit to the width of the sheath and the locations of the knots at $t = 2500$y. Similarly, an increase in $v_{sspray}(r_0)$ from 200 km s$^{-1}$ by 15% triggers disruptive thin-shell instabilities throughout the proboscis. Slower speeds result in unacceptable knot locations and average speeds. .
- The model does not account for asymmetries in the nebular brightness distribution on opposites side of the symmetry axis. (N.B.: only the surface brightness and not the nebular structure exhibit these asymmetries. So it appears that the uneven illumination of the nebula rather than an asymmetric distribution of internal pressures is the cause.)

6.3 Generalizations

*Spatial Structure*. Some of the striking features of M2–9 have counterparts in other pPNe or very young PNe. For example, like M2–9, Hen3–1475 has a pairs of inner bulbs and proboscises that converge to high-speed knots and fuzzy leading blobs with measurable proper motions. Hen2–320 has bulbs, a faint sheath, and feathers, like M2–9, but no knots or blobs. Mz3 has a bulb with a cylindrical extension that resembles the feathers of M2–9. Hen3–401, IRAS13208–6020, and IRAS17340–3757 are bipolar nebulae showing feathers on the outer edges of hollow cylinders, like M2–9, but with no knots, bulbs or sheaths in their interiors. CRL618, IRAS22036+5306, and OH231.8+04.2 each have pairs of lobes with knots at their tips but no bulb, sheath, or outer lobe. Two low-inclination bipolar pPNe, IRAS17150-3224 and perhaps Hen2-166, exhibit dusty plugs at the leading edges of their relatively wide lobes. The physical connections among these objects aren't clear. (Future hydro models would be very helpful to this end.)

The model presented here is obviously very specific to M2–9. Yet the processes that govern its early evolution may apply in some form to many other pPNe with feathers and coaxial pairs of lobes. In the earlier papers in this series, B+13 and B+17, we found that steady flows had to be light at the nozzle surface ($n_{flow}(r = r_0) < n_{amb}(r = r_0)$) in order for the lobes to be fashioned by the inertial pressure of its environment into the shapes that are observed. Light outflows are mandatory for creating lobes that appear hollow as they grow.

There are other interesting and general lessons to be learned about the density structures of pPNe from the hydro simulations. For example, the expansion speed at the tips of lobes formed by light stellar flows will always be less (and often substantially less) than the speed of injected gas as its speed is retarded by the displacement of ambient gas. Thus a general caution is in order: the speeds of the tips of hollow lobes will underestimate the injection speed at the nozzle. This applies to many well-observed pPNe such as M1–92, M2–56, Hen2–166, Hen3–401, IRAS 17150–3224, and Frosty Leo (Roberts 22).

Sprays (tapered conical flows) have the flexibility to fit a wide range of lobe shapes where as uniform-density cylindrical (parallel streamlines) and conical (diverging streamlines) flows do not. Specifically, steady, uniformly filled cylindrical jets can only create pole-shaped lobes similar to those of Hen3–401, IRAS13208-6020, and IRAS17340-3757. Additionally, forming a cylinder wider than a close stellar binary orbit (≈20 au) is tricky unless some sort of carefully shaped exterior collimator deflects diverging into parallel streamlines. Similarly, light conical flows will create leading arc-shaped plugs that act as pistons that displace ambient gas. The growing arcs must rapidly slow down as their area increases and they displace a wedge of downstream gas. Such a plug the plug is prone to highly destructive instabilities, as shown by simulations of spherical winds by Toalá & Arthur, 2014.

It is very awkward for steady flows of any geometry, whether uniform or tapered, to create pairs of on-axis knots that aren't the compressed remnants of plugs. N3/S3 are M2–9 are rare exceptions. The basic reason is knots are prone to crush unless they are much denser than the ambient gas (see B+17 for more about the crushing of clumps). The process of forming knot-like structures *in situ* was discussed by Akashi & Soker, 2008MNRAS.391.1063A and Akashi, et al., 2015MNRAS.453.2115A.

*Kinematics.* Earlier in this section we summarized our explanation for the seemingly linear rise of Doppler shift with distance seen in P-V diagrams in long-slit observations of M2–9 (sections 2 and 4.1; Fig. 1b). This explanation easily generalizes to the edges of all closed, expanding lobe sheaths of any width that are bounded by shocks.

Acknowledgements. It is a pleasure to thank Jonathan Carroll-Nellenback and the many people in the Physics-Astronomy Computational Group at the University of Rochester who developed and maintain the Astrobear code. Their help and support were essential. We gratefully acknowledge useful discussions with Noam Soker and Joel Kastner.

We gratefully acknowledge support for program AR-14563 that was provided by NASA through a grant from the Space Telescope Science Institute, which is operated by the Association of Universities for Research in Astronomy, Inc., under NASA contract NAS 5-26555. This paper is partially based on observations made with the NASA/ESA Hubble Space Telescope, obtained from the MAST Archive at the Space Telescope Science Institute, which is operated by the Association of Universities for Research in Astronomy, Inc., under NASA contract NAS 5-26555. Support for MAST for non-*HST* data is provided by the NASA Office of Space Science via grant NAG5-7584 and by other grants and contracts.

Facility HST (WFPC2 and NICMOS), Gemini (ALTAIR), and CTIO (Echelle)